\documentclass[iop,apj,revtex4]{emulateapj}
\usepackage{graphicx}
\usepackage{amsmath,amssymb}
\usepackage{tabularx}

\newcolumntype{L}[1]{>{\raggedright\arraybackslash}p{#1}}
\newcolumntype{C}[1]{>{\centering\arraybackslash}p{#1}}
\newcolumntype{R}[1]{>{\raggedleft\arraybackslash}p{#1}}
\newcolumntype{Y}{>{\centering\arraybackslash}X}

\usepackage{amssymb,amsmath}
\usepackage{txfonts}
\usepackage{xcolor}
\begin{document}

\title{von {K{\'a}rm{\'a}n-Howarth} equation for Hall
magnetohydrodynamics: Hybrid simulations}
%\title{Incompressible Hall-MHD exact turbulence law in collisionless plasmas:
%\\ Hybrid simulations}

% Turbulent cascade rates in collisionless plasmas: Incompressible Hall MHD
% exact law in 2D Hybrid simulations  

\author{Petr Hellinger\altaffilmark{1,2}}
\email{petr.hellinger@asu.cas.cz}
\author{Andrea Verdini\altaffilmark{3}}
\author{Simone Landi\altaffilmark{3,4}}
\author{Luca Franci\altaffilmark{3,5}}
\author{Lorenzo Matteini\altaffilmark{6}}

\altaffiltext{1}{Astronomical Institute, CAS,
Bocni II/1401,CZ-14100 Prague, Czech Republic}
\altaffiltext{2}{Institute of Atmospheric Physics, CAS,
Bocni II/1401, CZ-14100 Prague, Czech Republic}
\altaffiltext{3}{Dipartimento di Fisica e Astronomia, Universit\`a degli Studi di Firenze Largo E. Fermi 2, I-50125 Firenze, Italy}
\altaffiltext{4}{INAF -- Osservatorio Astrofisico di Arcetri, Largo E. Fermi 5, I-50125 Firenze, Italy}
\altaffiltext{5}{INFN -- Sezione di Firenze, Via G. Sansone 1, I-50019 Sesto F.no (Firenze), Italy}
\altaffiltext{6}{LESIA, Observatoire de Paris, Meudon, France}

\begin{abstract}
A dynamical vectorial equation for homogeneous incompressible Hall-MHD turbulence together with the exact scaling law for third-order correlation tensors, analogous to that for the incompressible MHD, is rederived and applied to the results of two-dimensional hybrid simulations of
plasma turbulence. At large (MHD) scales the simulations exhibits a clear inertial range where the MHD dynamic law is valid. In the sub-ion range the cascade continues via the Hall term but the dynamic law derived in the framework of incompressible Hall MHD equations is obtained only in a low plasma beta simulation. For a higher beta plasma the cascade rate decreases in the sub-ion range and the change becomes more pronounced 
as the plasma beta increases.  This break in the cascade flux can be ascribed to non thermal (kinetic) features or to others terms in the dynamical equation that are not included in the Hall-MHD incompressible approximation.
\end{abstract}


%\keywords{}

\maketitle

\section{Introduction}

Rarefied magnetized plasmas are ubiquitous in the astrophysical context.  There,
Coulomb collisions are so rare that non linear couplings driven by large scale medium
motions induce a turbulent cascade which reaches the kinetic scales of the
plasma constituent (ions and electrons) before collisional effects start to
dissipate its flux.  Turbulence in such collisionless/weakly collisional
plasmas is not well understood and remains one of the challenging problems of
astrophysics.

The magnetized plasma flow of the solar origin, the solar wind, constitutes a
natural laboratory for studying turbulence in collisionless plasmas
\citep{brca13}. In situ observations indeed show that the solar wind flow is
strongly turbulent, the magnetic and velocity fields exhibiting power-law
spectral properties as well as non-Gaussian statistical properties
\citep{mattal15d}. On relatively large scales the magnetic field power spectrum
of the observed time series is close to $f^{-5/3}$ reminding of the Kolmogorov
prediction for hydrodynamic (HD) turbulence. This spectrum is, however,
anisotropic with respect to the ambient magnetic field; along the magnetic
field direction the spectrum is steeper (close to $f^{-2}$) \citep{chen16}, in
agreement with theoretical expectations and numerical simulations based on the
magnetohydrodynamic (MHD) approximation.

Around the ion characteristic scales (proton gyroradius $\rho_i$ and proton
inertial length $d_i$, that are typically close to each other in the solar
wind)  the spectrum steepens.  This steepening is related to a change of the
physical behavior, in the sub-ion range the MHD approximation breaks and a more
accurate approximation including Hall and kinetic  effects is needed.  The
physical phenomena responsible for the steepening are not yet clearly
determined, many processes related to the dispersive (Hall) and  dissipative
(collisionless damping) phenomena ranges appear at similar scales 
\citep{mars06,alexal08}.  The
position of the transition from the large, MHD scales to the sub-ion scales (so
called ion spectral break) varies with the radial distance and depend on the
ion temperature (or on the ratio between the ion and magnetic pressures
$\beta_i$) \citep{brtr14,chenal14}.

Understanding of turbulence is strongly facilitated by existence of 
exact dynamical equations (which involve second and third order structures functions), obtained using a statistical approach assuming a homogeneity (and isotropy) of the medium.
The classical incompressible HD results  \citep{kaho38,kolm41b} has been extended to the incompressible MHD \citep{chan51,popo98a,popo98b,carbal09},
the incompressible Hall-MHD \citep{galt08}, and recently to the compressible Hall-MHD \citep{andral18}. 
In the context of the solar wind, effects of a homogeneous velocity shear \citep{wanal09} or an expansion \citep{hellal13} were also investigated.
Exact scaling laws, such as the well-known 4/3 and 4/5-laws, are obtained from these dynamical equations for the inertial range once a stationary state and infinite Reynolds number limit are assumed while still leaving a finite dissipative rate. 
The scaling predicted by the incompressible MHD exact laws were measured in the solar wind
and used to estimate the energy cascade rate (and the corresponding particle heating rate) in the solar wind\citep{sorral07,macbal08,marial08,stawal09,cobual15}.  
Beside the observations, numerical simulations constitute important means for investigating the properties of turbulence.  
The dynamical equations were tested in HD \citep{ishial09,gotoal02} and MHD simulations \citep{sorral02,mipo09,verdal15} in which a relatively good agreement with theoretical predictions is observed.
 However, due to the limited resolution, the range of scales where the scaling laws are verified is reduced (the scaling laws are exact only in the infinite Reynolds number limit).

\section{Incompressible Hall MHD}

For the transition between large MHD and sub-ion scales, numerical simulations
based on the hybrid approximation, where electrons are treated as a fluid
whereas ions are described fully kinetically, turned out to be very useful
\citep{paraal09,serval12,vasqal14,valeal14,serval15}.  Recent 2.5D hybrid simulations
\citep{franal15a,franal15b,franal16b} exhibit a clear double power-law behavior
of the magnetic field fluctuations in agreement with in-situ observations in
the solar wind. Such simulations are natural candidates for testing the exact
laws. As we are interested also in the sub-ion range, we want to look at the
incompressible Hall-MHD \cite[cf.,][]{pezzal17}. 

Assuming incompressibility, $\boldsymbol{\nabla}\cdot\boldsymbol{u}=0$, and a
constant plasma density for simplicity, $\rho=\mathrm{const.}$, taking the magnetic
field and the electric current in Alfv\'en units
$\boldsymbol{b}=\boldsymbol{B}/\sqrt{\mu_{0}\rho}$, and
$\boldsymbol{j}=\boldsymbol{J}/(e n)$, the Hall-MHD equation have this form:
\begin{align} \frac{\partial\boldsymbol{u}}{\partial
t}+(\boldsymbol{u}\cdot\boldsymbol{\nabla})\boldsymbol{u}-(\boldsymbol{b}\cdot\boldsymbol{\nabla})\boldsymbol{b}
& =-{\boldsymbol{\nabla}P}/{\rho}+\nu\mathrm{\Delta}\boldsymbol{u} \\
\frac{\partial\boldsymbol{b}}{\partial
t}\text{+}(\boldsymbol{u}\cdot\boldsymbol{\nabla})\boldsymbol{b}-(\boldsymbol{b}\cdot\boldsymbol{\nabla})\boldsymbol{u}&=(\boldsymbol{j}\cdot\boldsymbol{\nabla})\boldsymbol{b}-(\boldsymbol{b}\cdot\boldsymbol{\nabla})\boldsymbol{j}+\eta\Delta\boldsymbol{b};
\nonumber \end{align} here $P$ denotes the scalar total (particle and magnetic field)
pressure, and $\nu$ and $\eta$ denote the kinematic plasma viscosity and the
electric resistivity, respectively. 
Following  \cite{carbal09} we take the increment of the different quantities at $\boldsymbol{x}$ and
$\boldsymbol{x}^\prime= \boldsymbol{x} +\boldsymbol{l}$ ($\boldsymbol{l}$ being the spatial lag), $\delta
\boldsymbol{u}= \boldsymbol{u}(\boldsymbol{x}^\prime)-
\boldsymbol{u}(\boldsymbol{x})$,~\ldots{} and, assuming an isotropic turbulence
averaging over $\boldsymbol{x}$ (the averaging is denoted by $\langle\
\rangle$), we get the following equation for the magnetic, velocity, and total
second-order structure functions
$S_{b}=\langle\left|\delta\boldsymbol{b}\right|^{2}\rangle$,
$S_{u}=\langle\left|\delta\boldsymbol{u}\right|^{2}\rangle$, and
$S=S_{b}+S_{u}$ as functions of $\boldsymbol{l}$ 
\begin{align} \frac{\partial
S}{\partial
t}+\boldsymbol{\nabla}\cdot\left(\boldsymbol{Y}+\boldsymbol{H}\right)+A= &
-4\epsilon+2\nu\Delta S_{u}+2\eta\Delta S_{b} \label{yaglomH} 
\end{align} 
where the third order structure function $\boldsymbol{Y}$ is the MHD turbulent cascade flux \citep{carbal09,verdal15} 
\begin{equation} 
  \boldsymbol{Y}=\left\langle
  \delta\boldsymbol{u}\left|\delta\boldsymbol{u}\right|^{2}+\delta\boldsymbol{u}\left|\delta\boldsymbol{b}\right|^{2}-2\delta\boldsymbol{b}\left(\delta\boldsymbol{u}\cdot\delta\boldsymbol{b}\right)\right\rangle
  \label{Y}
\end{equation} 
and the third order structure function $\boldsymbol{H}$ is its Hall correction \cite[cf.,][for a different form of the Hall contribution]{galt08}
\begin{equation} 
\boldsymbol{H}=\left\langle 2 \delta\boldsymbol{b}\left(\delta\boldsymbol{b}\cdot\delta\boldsymbol{j}\right) - \delta\boldsymbol{j}\left|\delta\boldsymbol{b}\right|^{2}\right\rangle.
\label{H}
\end{equation} 
The last term at the l.h.s. of Eq.~(\ref{yaglomH}) $A=\left\langle \delta\boldsymbol{j}\cdot\delta\left[\left(\boldsymbol{b}\cdot\boldsymbol{\nabla}\right)\boldsymbol{b}\right]\right\rangle$, is a correction that we expect to be negligible in homogeneous plasma turbulence (this term is small in the present numerical simulations), and $\epsilon$ is the dissipation rate 
$\epsilon= \nu \left\langle \boldsymbol{\nabla} \boldsymbol{u} : \boldsymbol{\nabla} \boldsymbol{u} \right\rangle + \eta \left\langle \boldsymbol{\nabla} \boldsymbol{b} : \boldsymbol{\nabla} \boldsymbol{b}  \right\rangle $ 
(here `$:$' denotes the double contraction of two second-order tensors).  
 Eq.~(\ref{yaglomH}) is a dynamical equation that relates second and third order structure function generalizing the von~K\'arm\'an-Howarth equation in the framework of the incompressible Hall-MHD equations. 
An exact scaling law for a formally infinite-extending inertial range can be obtained assuming a stationary turbulent state 
in the infinite Reynolds number limit while retaining a finite dissipation rate,
$\boldsymbol{\nabla}\cdot(\boldsymbol{Y}+\boldsymbol{H})=-4\epsilon$, and, assuming isotropy,
one gets the scaling law 
 \cite[cf.,][]{carbal09} 
\begin{equation}
Y_r+H_r = -\frac{4}{3} \epsilon l
\label{YaglomScal} 
\end{equation}
where $l=|\boldsymbol{l}|$,  $Y_r$ and $H_r$ are the radial components (in the spherical coordinates corresponding to the lag space $\boldsymbol{l}$) of  $\boldsymbol{Y}$ and $\boldsymbol{H}$, respectively.
In a general anisotropic case of magnetized MHD the behavior is more complicated \cite[cf.,][]{verdal15}.

Note that the Hall term $\boldsymbol{H}$, Eq.~(\ref{H}), represents a correction to
the mixed terms in the MHD term $\boldsymbol{Y}$, Eq.~(\ref{Y}), replacing the
ion velocity by the electron one, $\boldsymbol{u}_e$, i.e.,
the magnetic field couples to the electron velocity field
\begin{equation} 
\boldsymbol{Y}+\boldsymbol{H}=\left\langle
\delta\boldsymbol{u}\left|\delta\boldsymbol{u}\right|^{2}+\delta\boldsymbol{u}_e\left|\delta\boldsymbol{b}\right|^{2}-2\delta\boldsymbol{b}\left(\delta\boldsymbol{u}_e\cdot\delta\boldsymbol{b}\right)\right\rangle
\label{Ye} 
\end{equation}
 as one may expect.

\section{Hybrid simulation results}
 Now we can directly test the prediction of  the dynamic
law, Eq.~(\ref{yaglomH}), in the hybrid simulation results where ions are
described by a particle-in-cell model whereas electrons are a massless, charge
neutralizing fluid \citep{matt94}.  The simulation setup is an extension of
previous simulations \citep{franal15a,franal15b,franal16b} and uses the 2D
version of the code 
Camelia ({http://terezka.asu.cas.cz/helinger/camelia.html}). 
The evolution of
initially isotropic protons with different values of $\beta_{i}$ ($1/16$,
$1/2$, and $4$) is numerically integrated in a 2D domain $(x,y)$ of size $256
d_i \times 256 d_i $ and resolution $\Delta x = \Delta y$. In order to
reduce the noise, a Gaussian smoothing on $3\times 3$ points is used
on the proton density and velocity in the code.
A uniform ambient
magnetic field $\boldsymbol{B_0}$, directed along $z$ and perpendicular to the
simulation domain is present whereas neutralizing electrons are assumed
isotropic and isothermal with $\beta_{e}=\beta_{i}$.  The system is perturbed
with an isotropic 2-D spectrum of modes with random phases, linear Alfv\'en
polarization ($\delta \boldsymbol{B} \perp \boldsymbol{B}_0$) and vanishing
correlation between magnetic field and velocity fluctuations.  These modes are
in the range  $0.02 \le k d_i \le 0.2$ and have a flat
one-dimensional power spectrum with rms fluctuations $\delta B$. The time step
$\Delta t$ for particles integration (the magnetic field is advanced with a
smaller time step $\Delta t_B = \Delta t/20$), the number of particle per cell
$N_{ppc}$, and the resistivity $\eta$, used to avoid energy accumulation at the
smallest scales, have been set specifically for each simulation and their
values are reported in Table~\ref{tabsim}.  We let the system evolve beyond the
time when the fluctuations of the parallel current maximize. This indicates a
presence of a well-developed turbulent cascade \citep{mipo09,serval15};
henceforth, we analyze properties of plasma turbulence at around these times 
($t_d=490\Omega_i^{-1}$, $350\Omega_i^{-1}$, and $290\Omega_i^{-1}$)
for each simulation.

\begin{table}[!t] 
\begin{tabularx}{0.5\textwidth}{c|YYYYYYYY} \hline \hline RUN
& $\beta_i$ &  $\delta B/B_0$ & $k_{\mathrm{inj}} d_i$ &   $\Delta x /d_i$ &
$\eta$ & $N_{ppc}$ &  $\Delta t$ \  $\Omega_i$ \\ \hline \hline 1  &$1/16$ &
$0.20$ & $0.2$  & $1/16$ & $2\; 10^{-4}$  & $1024$ & $0.0025$ \\ 2  & $1/2$ &
$0.25$ & $0.2$  & $1/16$ & $3 \; 10^{-4}$ & $4096$  & $0.01$ \\ 3  & $4$   &
$0.25$ & $0.2$  & $1/8$ & $5 \;  10^{-4}$  & $32768$  & $0.02$ \\ \hline
\hline 
\end{tabularx} 
\caption{List of Simulations and Their Relevant
Parameters ($\eta$ is given in the units of $\mu_0 v_{A}^2/\Omega_{i}$).}
\label{tabsim}
\end{table}

\begin{figure}[htb] 
\centerline{\includegraphics[width=7cm]{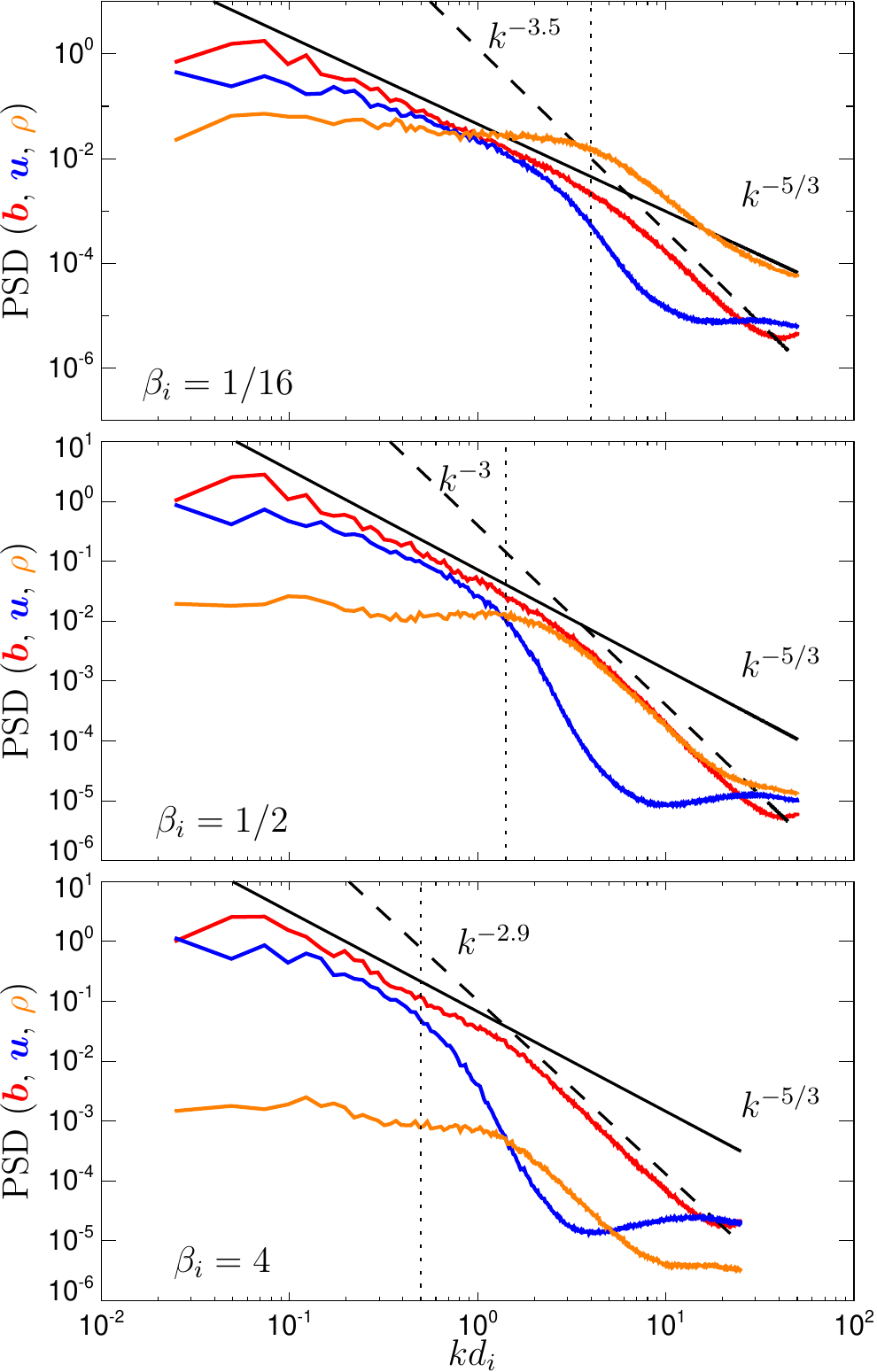}}
\caption{Power spectral densities of (red) the magnetic field, (blue) the
proton velocity field, and (orange) the proton density in the three simulations. 
The dotted lines denote $k\rho_i=1$.
 \label{specb}}
\end{figure}

Figure~\ref{specb} shows the power spectral density of the
magnetic field for the three simulations.  The simulated spectra exhibit two
power laws with a smooth transition at ion scales (the so called ion spectral
break), whose shape and position depend on the plasma beta: its scale
is close to $d_i$ for small betas whereas in high beta plasmas its around $\rho_i$ 
  \citep{franal16b}.
The magnetic power spectral slopes on large scales is $\sim -5/3$ whereas the
velocity power spectrum is less steep, closer to $-3/2$ than to $-5/3$.  In the
sub-ion range, as already observed in \citep{franal16b}, the magnetic power
spectrum steepens, with slopes about $-3.5$, $-3$, $-2.9$ in the three simulations.
The proton velocity fluctuating field decouples from
the magnetic fluctuations around the proton gyroscales. The sub-ion velocity
fluctuations have a limited scale range before reaching the noise level so that
it is difficult to distinguish between an exponential and a power-law
dependence.  Assuming the latter, as suggested by observations \citep{safral16}
and theoretical and numerical results \citep{meyral18}, the power spectrum of
the velocity fluctuations below the decoupling (above the noise level) is
compatible with steep slopes and spectral indices between $-5$ and $-6$.

Figure~\ref{specb} also shows the power spectral density
of the plasma density  $\rho$. The initial, transverse fluctuations
lead to formation of density fluctuation; the rms relative
density fluctuation $\delta \rho/\rho_0$ are $0.16$, $0.081$, and
 $0.017$ for the three runs going from low to high beta. As one
may expect the density fluctuations are stronger in a low beta plasma.

\begin{figure}[htb] 
\centerline{\includegraphics[width=7cm]{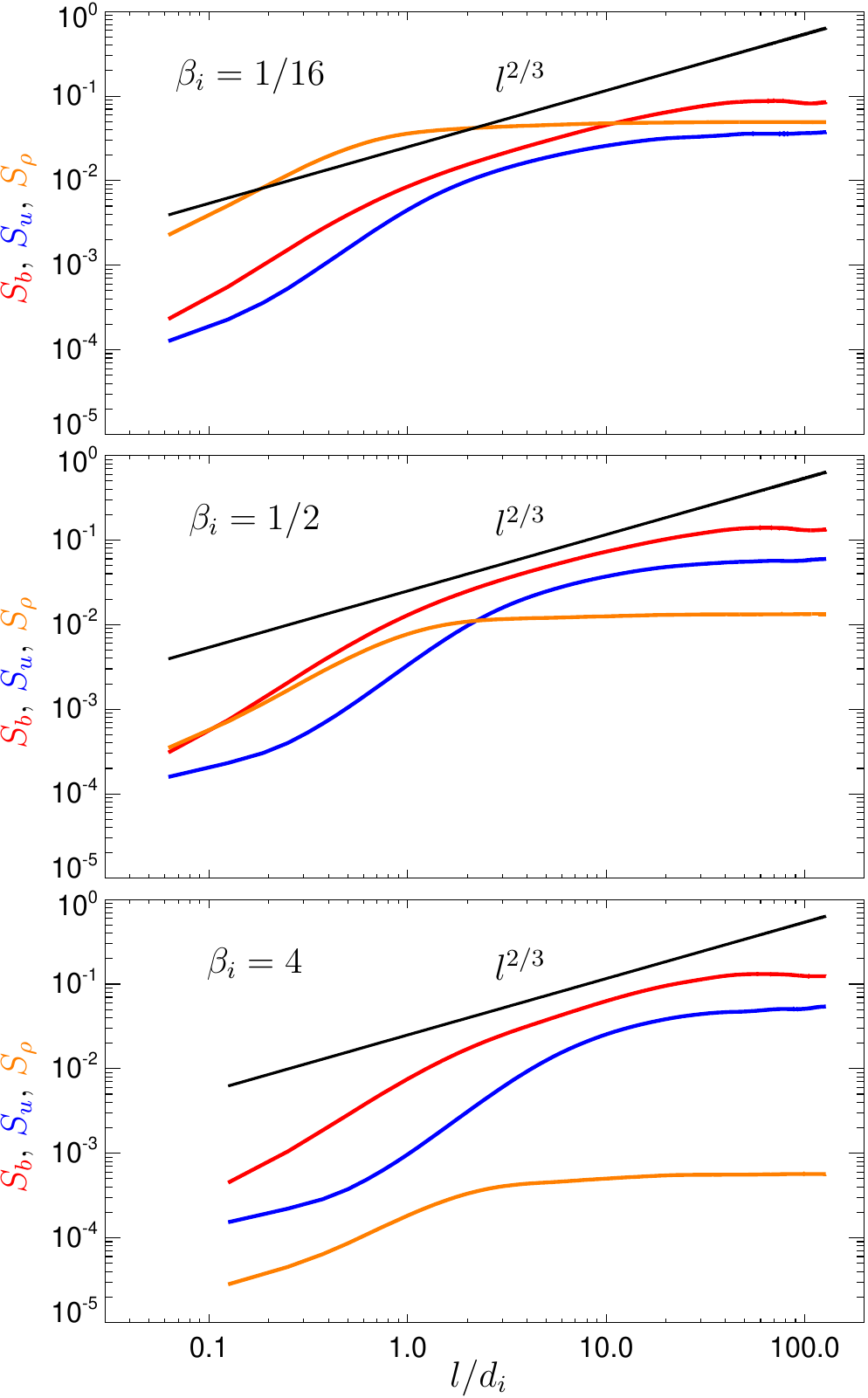}}
\caption{Second-order structure functions of (red) the magnetic field,
(blue) the proton velocity field, and (orange) the proton density in the three simulations.
\label{sf}} 
\end{figure} 

The spectral features observed in Figure~\ref{specb}
are partly reflected in the second-order structure functions of
the magnetic field $S_b$, the proton velocity $S_u$,
and the proton density $S_\rho$ (see Figure~\ref{sf}).
However, the properties of the inertial range, the spectral break, and
the sub-ion range are less clear compared to the power spectra.

\begin{figure}[htb] 
\centerline{
\includegraphics[width=7cm]{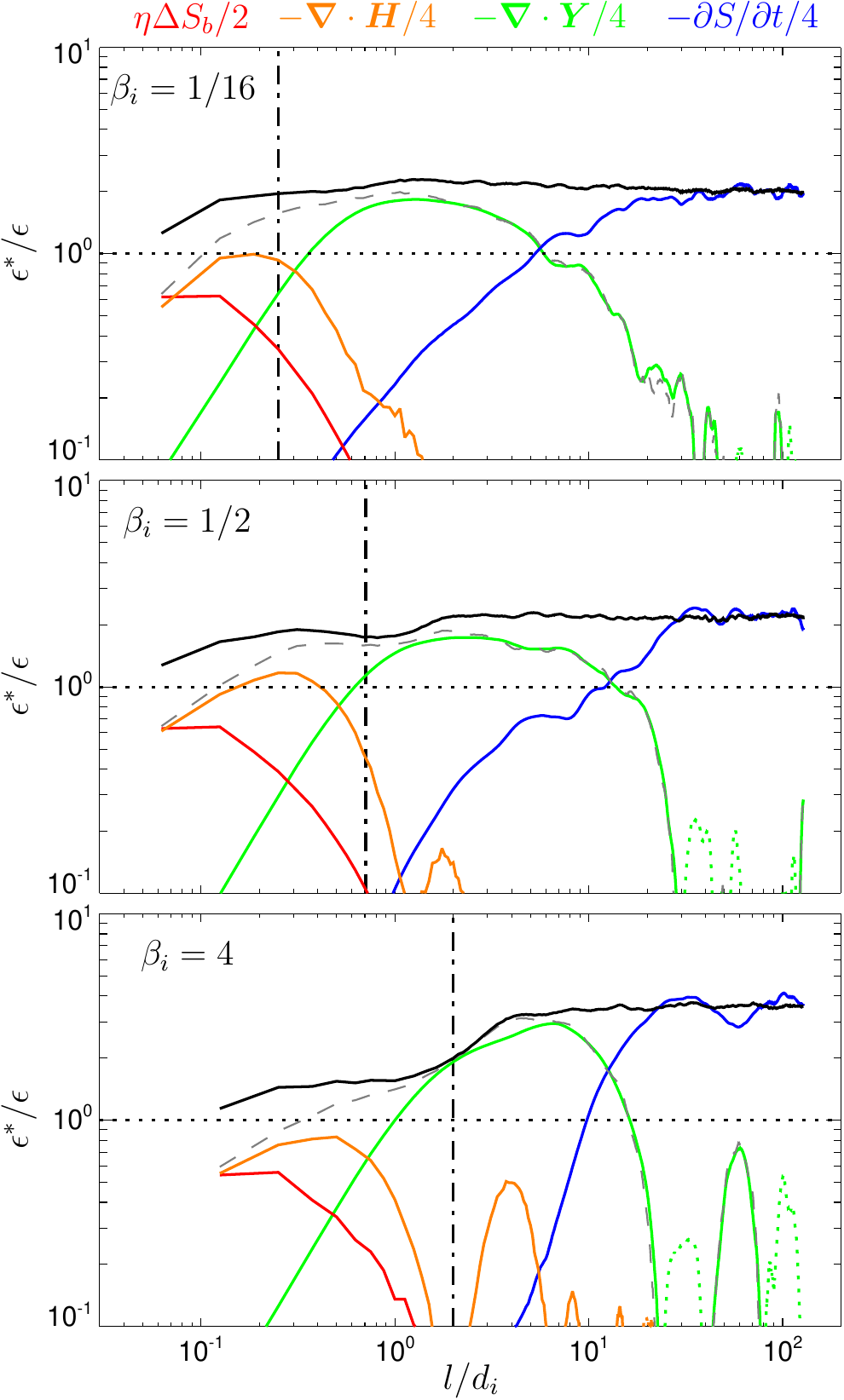}
}
\caption{Exact law in hybrid simulations $\beta_i=1/16$, $\beta_i=1/2$, and
$\beta_i=4$: The cascade rate $\epsilon^*$ normalized to the resistive heating
rate $\epsilon$ as a function of $l$ is shown as a black curve. The different
contributing terms to $\epsilon^*$ are also shown as: (blue) $-{\partial
S}/{\partial t}/4$, (green) $-\boldsymbol{\nabla}\cdot\boldsymbol{Y}/4$,
(orange) $-\boldsymbol{\nabla}\cdot\boldsymbol{H}/4$, and (red) $\eta\Delta
S_{b}/2$ (note that dotted lines denote negative values). The thin gray dashed
lines display the sum $-\boldsymbol{\nabla}\cdot(\boldsymbol{Y}+\boldsymbol{H})/4$.
  The dash-dotted
lines denote $l=\rho_i$. \label{yag}} 
\end{figure}

For the test of the dynamical equation, Eq.~(\ref{yaglomH}), we define the cascade rate $\epsilon^*$ as 
\begin{equation} 
\epsilon^*= -\frac{1}{4}\frac{\partial S}{\partial t}-\frac{1}{4}\boldsymbol{\nabla}\cdot\left(\boldsymbol{Y}+\boldsymbol{H}\right)+\frac{1}{2}\eta\Delta S_{b} 
\end{equation}
Figure~\ref{yag} shows the test of the dynamic law, the cascade rate $\epsilon^*$ normalized to the resistive heating rate $\epsilon$ with the different contributing terms to $\epsilon^*$.
Note that in these simulations the correcting term $A$ is not important, $|A|/4 \lesssim 0.1 \epsilon$ (not shown).
The structure functions (and $\epsilon$)  are calculated at two times ($t_d+\delta t$ and $t_d$ separated by $\delta t=10\Omega_{i}^{-1}$)  around the maximum turbulence activity and averaged; $\partial S/\partial t$ is estimated by $(S(t_d+\delta t)-S(t_d))/\delta t$.

In all simulations we observe that at MHD scales  $\epsilon^*\sim-\partial S/\partial t/4-\boldsymbol{\nabla}\cdot\boldsymbol{Y}$.
The term $\partial S/\partial t$ is due to the energy decay at injection scales given by the (decaying) initial condition whereas the region, where $\boldsymbol{\nabla}\cdot\boldsymbol{Y}$ dominates, covers the range of scales where a Kolomogorov-like spectrum is observed, Fig.~\ref{sf}, and represents the inertial MHD range. There, $\boldsymbol{\nabla}\cdot\boldsymbol{Y}$ varies only weakly and
the radial component $Y_r$ (in the cylindrical coordinate system corresponding to  $\boldsymbol{l}$) is roughly proportional to $l$ (not shown here), representing an equivalent of the hydrodynamic exact scaling law.

Crossing $l\sim d_{i}$ the Hall term flux $\boldsymbol{\nabla}\cdot \boldsymbol{H}$ starts to grow and becomes the dominant one at sub-ion scales ($\lesssim d_i$), although this term is important only in a narrow range of scales and no linear scaling of $H_r$ on $l$ is observed. The sum of the two inertial contributions, $\boldsymbol{\nabla}\cdot(\boldsymbol{Y}+\boldsymbol{H})$ (the thin gray dashed lines in Fig.~\ref{yag}) shows that the reduction in the MHD flux is partially compensated by the Hall flux and, where $\boldsymbol{\nabla}\cdot(\boldsymbol{Y}+\boldsymbol{H})$ is about constant, $Y_r+H_r$ is roughly proportional to $l$. This is especially true in the low $\beta$ simulation while a fraction of the flux is more and more lost as $\beta$ increases. The position of the  Larmor radius $\rho_i$ for each simulation (the vertical dot-dashed lines) suggests some kind of correlation between the decreasing inertial flux and $\rho_{i}$. 
 
Only at the smallest scales the magnetic diffusion term, $\eta\Delta S_{b}/2$, contributes significantly to the total cascade rate. In
the low $\beta$ simulations the sum of all contributions (black line) is quite constant at all scales. For the $\beta_{i}=1/2$ a small reduction of the total
cascade rate is observed near $l\sim d_{i}\sim\rho_i$ and such the reduction becomes significant in the high $\beta$ simulation (around $l\sim \rho_i$). In all cases,
however, $\epsilon^*$ at the injection-MHD scales is more than twice the Joule dissipation rate $\epsilon$. 

Even if we use a large number of particle per cell, a part of the smallest scales is influenced by the particle's noise, especially in the ion velocity field (the magnetic field is much less affected \citep{franal15b}). However, ion velocity does not contribute to the Hall flux, $\boldsymbol{H}$, and comparison of two simulations with a different number of $N_{ppc}$ (not shown here) confirms that sub-ion cascade rate is only weakly affected by the noise level.  
The resistivity used in the hybrid simulations is a free parameter that is used to avoid accumulation of the cascading energy on small scales and in some respects replace the full electron physics. Increasing the resistivity leads typically to a reduction of the Hall cascade rate because dissipative scales shift toward larger scales and overlap with the Hall term but $\epsilon^*$ is not significantly modified.

\section{Discussions and Conclusions}
In this work we have tested by means of hybrid simulations
a dynamical vectorial law, Eq.~(\ref{yaglomH}), which is the generalization of the von K\'arm\'an-Howarth equation for incompressible hydrodynamic turbulence in the framework of incompressible Hall MHD equations.
Simulations show that in the low $\beta$ regime Eq.~(\ref{yaglomH}) is reasonably well satisfied and 
an indication of the MHD inertial regime (where $\boldsymbol{\nabla}\cdot\boldsymbol{Y}\simeq {\mathrm const}$) is obtained. At sub-ion scales the decreasing of the MHD flux is compensated by an increase of the Hall term, $\boldsymbol{\nabla}\cdot\boldsymbol{H}$, which extends the inertial regime below the scale of the spectral break. Increasing the plasma $\beta$ the Hall flux only partially compensates the reduction of the MHD one, meaning that in the intermediate and high $\beta$ regimes Eq.~(\ref{yaglomH}) breaks. 

The break of Eq.(\ref{yaglomH}) at high $\beta$ does not seem to be related to compressibility effects. Indeed, ratios between density and magnetic field power spectra reveal that compressible effects are less important as $\beta_i$ increases both at MHD and kinetic scales. 
However, the incompressible description of the Hall MHD equations from which the dynamic law is derived does not take into account the other relevant quantity at ions scales, the ion-Larmor radius. For instance, this scale seems to be relevant in determining the break at ions scales in high beta plasmas \citep{chenal14,franal16b} and characterises the polarisation properties of turbulence mediated by kinetic-Alfv\'en-wave-like fluctuations  \citep[e.g.,][]{scheal09,boldal13,chenal13b,franal15b}.
Alternatively, the reduction of the cascade rate $\epsilon^*$  could be related to non-thermal features playing at the ion gyroradius also not retained in the Hall-MHD approximation with the scalar pressure. Indeed, the work of \cite{delsal16}   indicates that pressure anisotropies/nongyrotropies appear at around  the scale of $\rho_{i}$.   
In the low beta case $\rho_{i}$, is significantly smaller then $d_{i}$ and at scale near the dissipative ones. As $\beta$ increases, $\rho_{i}$ becomes comparable to $d_{i}$ and a larger portion of the ion velocity distribution function can interact with turbulent fluctuations. Consequently, the reduction of $\epsilon^*$ moves toward larger scales and it is more pronounced.

Another difference between the kinetic simulations and the predicted cascade rate is its value. 
The cascade rate $\epsilon^*$ is typically a factor two larger then resistive losses $\epsilon$.
One possibility is that a non-negligible fraction of the dissipation is carried by numerical effects.
The dissipation rate $\epsilon$ may be underestimated due to the 2-nd order numerical scheme which introduces an effective dissipation $\propto \nabla^2$ in the magnetic field. If it is so, $\epsilon$ would be greater while $\epsilon^*$ remains unchanged (except at the smallest scales) thus reducing the discrepancy between the two.
The other possibility is that the energy cascading from MHD scales is partially transferred to ions via physical processes beyond the Hall-MHD approximation, possibly connected with non-thermal features as suggested by \cite{yangal17}. 
The ion heating rate in our simulations is indeed comparable to $\epsilon^*-\epsilon$; however, for a detailed study of the relation between $\epsilon^*$ and the ion (and electron) heating rates one needs to include the full electron kinetics.

These results are very robust, since we observed similar behavior in all our our previous
2D hybrid simulations \citep{franal16b,cerral17}; these simulations further indicate 
that the decrease of $\epsilon^*$ in the sub-ion region evolves continuously from low to high beta plasma.
The present work supports the validity of
the estimates of the cascade rate in the solar wind \citep{sorral07,macbal08,stawal09,marial11,cobual15}.
However, it is necessary to include compressible effects \citep{baga13,hadial17,andral18} as well as
the effects of the anisotropy of the cascade in the full 3D geometry \citep{verdal15} in future work.

It is interesting to note that perpendicular spectral properties of 3D hybrid simulations are
in many respects similar to their 2D counterparts \citep{franal18}; we expect an analogous behavior
in the case of the exact law.
The present work also needs to be extended to full particle simulations to compare the cascade rate determined from the
exact law with the actual heating rates of the different particle species \citep{wual13,mattal16}.
In concluding, the present work suggests that the transition from the
MHD to sub-ion scales is governed by a combination of an onset of Hall
cascade (with the characteristic scale $d_i$) and a formation of ion
non-thermal features related to dissipation (with the characteristic scale $\rho_i$).

\acknowledgments
The authors thank T.~Tullio for a continuous enriching support and
acknowledge useful discussions with W. H. Matthaeus and L. Sorriso-Valvo.
P.H. acknowledges grant 18-08861S of the Czech Science Foundation.
The authors acknowledge CINECA for HPC resources under the ISCRA initiative (grants HP10BUUOJM, HP10BEANCY, HP10B2DRR4,
HP10CGW8SW, HP10C04BTP).
The (reduced) simulation data are available at the Virtual Mission Laboratory Portal (http://vilma.asu.cas.cz).

\end{document}